# Turing Pattern Engineering Enables Kinetically Ultrastable yet Ductile Metallic Glasses

Huanrong Liu[1], Qingan Li[1], Shan Zhang[2,3], Rui Su[4,*], Yunjiang Wang[5,6,*]，Pengfei Guan[2,3,1,*]

[1] Beijing Computational Science Research Center, Beijing 100193, China.

[2] Advanced Interdisciplinary Science Research (AiR) Center, Ningbo Institute of Materials Technology and Engineering, Chinese Academy of Sciences, Ningbo 315201, China.

[3] State Key Laboratory of Advanced Marine Materials, Ningbo Institute of Materials Technology and Engineering, Chinese Academy of Sciences, Ningbo 315201, China.

[4] College of Materials & Environmental Engineering, Hangzhou Dianzi University, Hangzhou 310018, China.

[5] State Key Laboratory of Nonlinear Mechanics, Institute of Mechanics, Chinese Academy of Sciences, Beijing 100190, China.

[6] School of Engineering Science, University of Chinese Academy of Sciences, Beijing 101408, China

*Corresponding authors. surui@hdu.edu.cn (R.S.), yjwang@imech.ac.cn (Y.W.), pguan@nimte.ac.cn (P.G.)

**Abstract**

Enhancing the kinetic stability of glasses often necessitates deepening thermodynamic stability, which typically compromises ductility due to increased structural rigidity. Decoupling these properties remains a critical challenge for functional applications. Here, we demonstrate that pattern engineering in metallic glasses (MGs) enables unprecedented kinetic ultrastability while retaining thermodynamic metastability and intrinsic plasticity. Through atomistic simulations guided by machine-learning interatomic potentials and replica exchange molecular dynamics, we reveal that clustering oxygen contents, driven by reaction-diffusion-coupled pattern dynamics, act as localized pinning sites. These motifs drastically slow structural relaxation—yielding kinetic stability comparable to crystal-like ultrastable glasses, yet retaining an energetic as-casted state. Remarkably, the thermodynamically metastable state preserves heterogeneous atomic mobility, allowing strain delocalization under mechanical stress. By tailoring oxygen modulation via geometric patterning, we achieve a ~200 K increase in the onset temperature of glass transition ($T_{onset}$) while maintaining fracture toughness akin to conventional MGs. This work establishes a paradigm of *kinetic stabilization without thermodynamic compromise*, offering a roadmap to additively manufacture bulk amorphous materials with combined hyperstability and plasticity.

**Key words :** metallic glasses, kinetic stability, Turing pattern, reaction-diffusion model

**Introduction**

Stability of disordered materials has been a long-term pursuit, tracing back to the concept of ideal glassy state predicted by Kauzmann[1]. Unlike ordinary as-cast glasses, most accessible ultrastable glasses exhibit a heat capacity as low as their crystalline counterparts, as demonstrated in amorphous silicon[2] and indomethacin[3]. However, the thermodynamically crystal-like stability have been intertwined with the challenge of balancing kinetic stability and mechanical ductility[4,5]. Conventional strategies, such as physical vapor deposition for fabrications[6,7] and computational techniques like particle swapping[8], often enhance both kinetic and thermodynamic stability by driving systems into deeper energy basins in the potential energy landscape (PEL). Yet, these approaches inevitably amplify structural rigidity, leading to increased brittleness—a critical limitation for practical applications where plasticity is paramount. For instance, vapor-deposited ultrastable glasses exhibit exceptional thermal stability but are constrained to thin-film geometries and anisotropic structures[9], while randomly particle pinning disrupts translational invariance[10,11], altering mechanical responses and relaxation pathways. These inherent trade-offs underscore a fundamental dilemma. That is, the enhancement of kinetic stability in glasses is typically coupled with thermodynamic stabilization, which sacrifices ductility and undermines applicability. A pivotal question thus emerges: *Can kinetic stability be decoupled from thermodynamic stabilization to preserve plasticity?*

The clue to this question stems from the vapor-deposited ZrCuAl metallic glasses (MGs)[12,13], which exhibit enhanced kinetic stability without any thermal compromise. The anomaly is explicitly attributed to the unique interaction in metallic environments. On the one hand, metallic bonding naturally outperforms others in the proliferation of basins in whether strong or fragile glassy PELs, as demonstrated by the prolonged aging experiments of Ce-, Zr-, Al-, Ni-based MGs at room temperature[14,15]. On the other hand, the itinerant electrons[16,17] of 3-$d$ and/or 4-$f$ orbitals provide additional protections for barrier profiles, enabling additional kinetic stability. Recent investigations into doped Zr-based MGs have also revealed exceptional kinetic[18] and mechanical[19] stability simultaneously, offering another compelling perspective to the underlying impurity impacts on bonding-assisted stabilization of MGs. However, the formation mechanism responsible for this intrinsic stability is currently not well understood, posing an obstacle to the subsequent regulatory modification.

Here, we propose a novel paradigm for designing MGs with decoupled stability and ductility through spatial modulation of oxygen dopants. By leveraging oxygen's strong affinity for zirconium and its sluggish diffusion dynamics, we engineer oxidation-induced pinned structures (OPSs) with tailored spatial patterns. Unlike conventional doping strategies that homogenize solute distributions[20], our approach utilizes reaction-diffusion principles to create *thermodynamically metastable yet kinetically ultrastable* configurations. Atomistic simulations, supported by machine-learning interatomic potentials (MLIPs), reveal that patterned oxygen-rich regions act as localized constraints, drastically slowing structural relaxation while maintaining a heterogeneous energy landscape. Crucially, the thermodynamic metastability of these configurations preserves the atomic mobility necessary for plastic deformation, avoiding the brittleness associated with deeply equilibrated ultrastable states. By decoding the configuration-governed slowdown mechanisms and density-dependent activator-inhibitor transitions, we establish a stability-oriented framework, bridging theoretical models with practical design strategies for ultrastable glasses in chemically complex amorphous systems.

## 2. Methods

### 2.1 Molecular dynamics sampling

We first equilibrated a plate-like, oxygen-free $Cu_{50}Zr_{50}$ sample (11.5×11.5×1.4 nm³, containing 16,384 atoms) at 1500 K. The sample was then quenched to 1 K in an isothermal-isobaric ensemble (as proposed by Parrinello and Rahman[21]) with zero external pressure. In our work, all cooling and heating rates are consistently maintained at $10^{12}$ K/s except the preparation of oxygen-free metallic glasses with the cooling rate varying from $10^{11}$ to $10^{14}$ Ks$^{-1}$. Structural optimizations by the conjugate gradient method were afterwards conducted to explore corresponding inherent structures.

For O-doped samples, we homogenously inserted oxygen atoms ($\overline{C_O} = 0.4\%, 0.9\%, 1.8\%, 3.0\%, and\ 4.7\%$) at atomic-Voronoi vertices of the O-free reference. Replica exchange molecular dynamics (REMD) was performed to enhance sampling efficiency of oxidation-induced pinned structures (OPS). Tempering sets in four stages from a target temperature at 1000 K, to intermediate and high temperatures at 1500 K, 2000 K, and 2500 K. A temperature swap between adjacent ensembles will take place in every 50 ps. Each attempted swap is either accepted or rejected based on a Metropolis criterion $P(1 \leftrightarrow 2) = \min(1, \exp[(\frac{1}{k_B T_1} - \frac{1}{k_B T_2})(U_1 - U_2)])$. A 3-ns trajectory at the target temperature was generated from a hybrid ensemble of canonical MD and REMD. Noted that synchronization in REMD is determined by technical considerations such as computational cost, rather than thermodynamic processes such as aging and crystallization[22]. Namely, REMD is an augmented sampler to broaden the energy spectrum of pinned structures at the cost of temporal information.

The OPS was topologically constructed by the "-Zr-O-Zr-" unit and a 2.8 Å cutoff was applied to select nearest neighbors. Subsequently, a point pattern with locally convex metrics can be built by the triangular tessellation. The region within such a smooth hull, visualized by yellow surface meshes, is referred to as an OPS.

## 2.2 Zr-Cu-O neural network potentials

All atomistic simulations in this work were conducted on LAMMPS packages[23] in conjunction with our newly developed Zr-Cu-O neural network potentials. Model parameters and corresponding raw datasets were generated and upgraded by an iterative pipeline workflow. The concurrent learning framework can significantly reduce training costs of multi-component MG systems. More details about the training can be found in the published note of ref. [19].

## 2.3 Evaluation of kinetic, mechanical and dynamical properties

A stepwise heating protocol was applied to evaluate the $T_{\text{onset}}$. To refine the accuracy of transition indicators, the interval of temperature hopping was adaptively rescaled to 10K around critical points by conducting isothermal stays at parallel increments. We fit the enthalpy curve above and below the transition with two separate polynomials and ensure continuity in the first and second derivatives by introducing sigmoid smoothing. The maximum linear approximation of the second derivatives is used to determine the glass transition point through its intercept[24]. Kinetic glass transition temperature rather than thermal fictive temperature is investigated in our simulations due to the ultrahigh melting point of ceramic-like oxides[25], which hinders determination of the liquid reference in the Moynihan area matching method.

Mechanical simulations were performed using a dynamical loading protocol with a strain rate of $1 \times 10^7 s^{-1}$, which approaches the time duration limit of simulations driven by MLIPs. Lees-Edwards boundary conditions were used throughout for the generation of translation invariant shear flow. The samples for mechanical tests were prepared in cubic boxes (11.5×11.5×11.5 nm³, containing 131,072 atoms). Simulations were conducted on 20 replicas, where the initial momenta adhere to the Maxwell distribution. Plasticity is characterized by the magnitude of stress drop, which is also associated with the strain localization parameter[26] $\psi = \sqrt{\frac{1}{N}\sum_{i=1}^{N}(\eta_i^{Mises} - \eta_{ave}^{Mises})^2}$, where $\eta_{ave}^{Mises}$ is the average von Mises strain in the loading snapshot.

To understand the microscopic origin of kinetic anomaly, we explicitly evaluated dynamical properties by stratification. The distances between oxygen atoms and the target atom were calculated by the minimum image convention and the shortest was selected as a label for the stratification. The configurations were then divided into slices with thickness of 0.75 Å. Only slices with enough atoms (> 100) were collected for calculations. In details, we generalized the conventional form of incoherent scattering functions as $F_{s,d}(\boldsymbol{q}, t) = \frac{1}{N}\sum_{j=1}^{N} \exp\left[i\boldsymbol{q} \cdot (\boldsymbol{r}_j(t) - \boldsymbol{r}_j(0))\right] \delta[d_j(0) - d]$, where $d_j(0)$ represents the distance of atom j from the OPS at $t = 0$. The first peak of structural factors ($q = 2.71 \text{ Å}^{-1}$) was applied to replace wave vectors in the real part of calculations (see SI Figure 1).

**2.4 Reaction-diffusion model implementation**

The models were implemented by temporal and spatial discretization with dimensionless $\Delta t = 0.005$ and $\Delta L = 0.5$ on a grid of $1024 \times 1024$. Periodic boundaries were used in both sides to minimize the edge effect. The rate of density change was calculated by forward differences with 90000 timesteps. The two-dimensional second derivative operator was discretized by a numerical central-difference kernel. Spatial average of global density was specified in three values ($\langle \tau \rangle =$ 3.0, 4.5 and 6.0 ) to mimic different doping concentrations. To specify the initial concentration of slow and fast particles, we performed canonical MD simulations for 400 ps (enough for atoms to escape from cages) on a homogeneously doped $Cu_{50}Zr_{50}$ sample and its oxygen-free reference. The mean squared displacement of reference was chosen as a proxy of the classifier. Specifically, in our model, particles slower than their constraint-free counterparts act as activators $u$, while the remaining faster-moving atoms function as inhibitors $v$. Meanwhile, the diffusion ratio δ was estimated by a generalized diffusion coefficient $D_X = \frac{1}{6N_X t}\sum_{j=1}^{N_X}\Delta r_j^2(t)$, where $X = u, v$. The parameters in reaction terms of Equations 1a and 1b were determined in the pattern-allowed spinodal region ($\alpha = 4$ and $\beta = 1$). To trigger the fluctuation for pattern formation, a slight random perturbance was carried out on the initial distribution of activators and inhibitors.

**2.5 Pattern mapping and patch-size distribution**

Pattern masks were prepared in a snapshot predicted by Equations 1a and 1b. Pixels on masks were binarized by a non-shrinking criterion on activator ($u > \langle u_- \rangle = 4.37$). Binary pixels were grouped into a cluster if they shared the same Von Neumann neighborhood. We also prepared a plate-like reference configuration in the size of $13.2 \times 13.2 \times 1.6$ nm³. Voronoi tessellation was performed based on VORO++ packages[27] to build a candidate library of vertices (see SI Figure 2). Next, the concentration filter assigns the number of selected vertices based on the corresponding cluster size and optimizes the embedding positions by randomly swapping nearest-neighbor Cu atoms with next-nearest-neighbor Zr atoms, ensuring the coordination bias of Zr-O pairs. Structural optimizations were carried out to conform with MD-sampled OPSs.

The patch radius was calculated using a coarse-grained approach, based on the reconstruction of atomic profiles into pattern masks. We implemented this reverse mapping method by calculations of local density $u$ of slow particles. Pixels were then binarized and grouped using the same criteria as the pattern-mapping method. Finally, we calculated the dimensionless radius by $\bar{r} = \frac{r}{r_0} = \sqrt{N_c}$ ($N_c$ represents the occupied number of grids in the cluster) and probability density $g(\bar{r})$. In our model, $r_0 = 5\,\text{Å}$. The theoretical prediction from the LSW model writes $g_{\text{LSW}}(\bar{r}) = \frac{4\bar{r}^2}{9\mu^3}(\frac{3\mu}{3\mu+\bar{r}})^{7/3}(\frac{3\mu}{3\mu-2\bar{r}})^{11/3}e^{\frac{2\bar{r}}{2\bar{r}-3\mu}}$. For $\bar{r} \geq 3/2\mu$, $g_{\text{LSW}}(\bar{r}) = 0$. Noted that the maximum likelihood estimates instead of numerical averages of $\mu$ were used, leading to a left skewed distribution. The value of $\mu$ can be determined by the equation $3n + \frac{7}{3}\sum_{i=1}^{n}\frac{3\mu}{3\mu+\bar{r}_i} - \frac{11}{3}\sum_{i=1}^{n}\frac{3\mu}{3\mu-2\bar{r}_i} + 3\mu\sum_{i=1}^{n}\frac{2\bar{r}_i}{(2\bar{r}_i-3\mu)^2} = 0$, where $n$ is the number of patches.

## 3. Results

### 3.1 Oxidation-induced pinning patterns in Zr-based metallic glasses

Conceptually, light-element doping serves as a third pathway towards hyperstable glasses besides randomly bonding and swaping in simulations, the former of which is further relevant to practice of material engineering. Figure 1a clearly depicts the slowdown of glassy dynamics due to formation of the motif of oxidation-induced pinned structures (OPSs). The as-sampled plate-like structure ($\overline{C_O}\sim 4.7$ at. %) exhibits a discernible decrease in mobility, attributed to an infinitesimal diffusion coefficient of oxygen atoms at 1000 K. Similar self-trapped behaviors have been observed in the O-doped $Cu_{40}Zr_{60}$ (at. %) glass-forming liquid[28] and $Cu_{70}Zr_{30}$ (at. %) nanomembranes[18]. Drawing inspiration from the concept of random pinning in equilibrium, we adopted a more intricate, yet realistic protocol designed to elucidate the transitions between pinned structures, as illustrated in Figure 1b. Topographic view on the PEL indicates proliferation of deep basins by oxygen addition, in consistent with prior knowledge[29-31]. The dynamic slowdown via oxygen pinning based on machine-learning interatomic potential (MLIP) is consistent with the seminal work[11] which evidences that frozen structures generally constrain the available configurational space of fluid particles. Additionally, aggregation of isolated oxygen contents is allowed to form the O-centered clusters[19,28,32]. All as-sampled OPSs will be sorted by corresponding energy states. We implemented the modeling protocol in a temperature-swapping framework, which is known as replica exchange molecular dynamics (REMD)[33], ensuring an efficient sampling in multi-canonical ensembles (see **Methods 2.1** and Figure 1c). As shown in Figure 1d, the ability to escape constraints was quantitatively encoded at the atomic level using the displacement factor $q(\Delta r) = 1 - \min(\overline{\Delta r}, \Delta r)/\overline{\Delta r}$, where $\overline{\Delta r}$ represents the mean displacement in canonical MD after isoconfigurational sampling for 2 ns. REMD can significantly lower atomic correlations and effectively explore configurational spaces in a manageable time, even for the pinned structures which exhibit strong localization in conventional MD. Figure 1e illustrates the evolution on energy and morphology while utilizing hybrid processing protocols to explore deeper configurational space in the PEL with pinned structures. As shown in snapshots, distinct OPSs are accessible by tuning energy states.

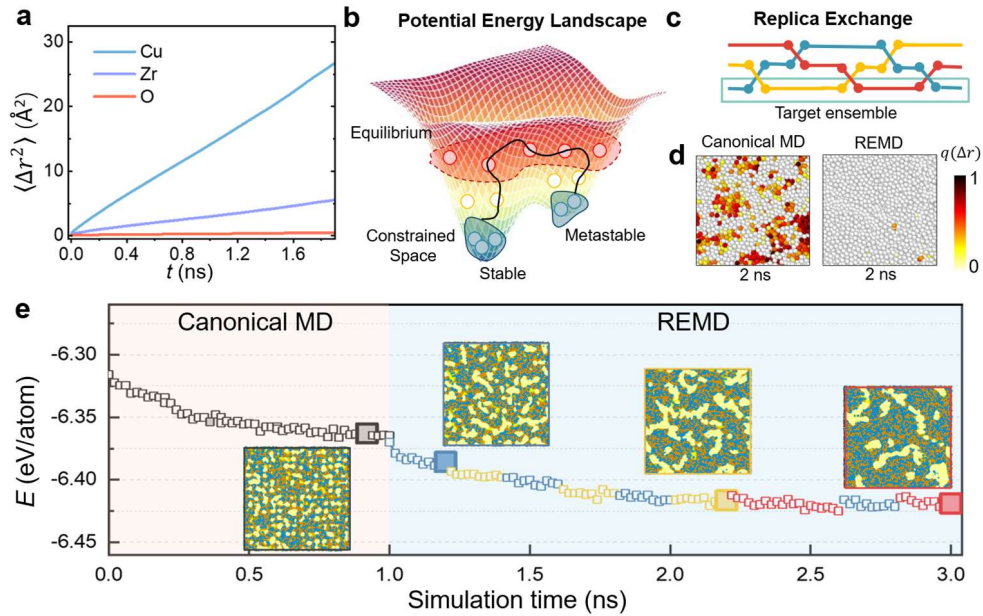

**Figure 1 Efficient exploration of the oxidation-induced pinned structures in metallic glass. a.** Elemental contributions to mobility with a bathing temperature of 1000 K. **b.** Sketch map of the potential energy landscape illustrating the oxidation-induced pinned structures. Constrained configurational spaces by pinning are highlighted by blue dashed circles. The black solid line represents a pathway connecting stable and metastable states, which are separated by a ratchet-like energy barrier. **c.** Schematics of the molecular dynamics with replica exchange (REMD). Sampling pieces from different protocols are concatenated to reproduce the target canonical ensemble, marked by the dashed blue box. **d.** Atomic color coding of the residual correlation after a simulation time of 2 ns, in the canonical MD- and REMD-sampled specimen respectively. **e.** The evolution of morphologies of oxygen species during the hybrid process. Four characteristic states are highlighted by enlarged squares along the pathway leading to energy-favorable basins. Spatiotemporal snapshots of the pinned structures in metallic glasses are also presented (orange spheres, Cu atoms; blue spheres, Zr atoms; yellow spheres and regions, pinned O atoms and clustering networks).

## 3.2 Simultaneous enhancement of kinetic stability and plasticity in metastable patterns

The efficient methodology of REMD combined with MLIP facilitates investigations on thermodynamic and kinetic spectrums of O-doped metallic glasses with quantum mechanics accuracy. Given the fact that OPSs are significantly dependent on thermodynamic state in supercooled regime[34], our simulations were all performed from the inherent structures (IS) where oxygen contents are fully frozen. Samples with notable difference in the energy basin $E_{IS}$ were chosen for comparison. We applied a step-by-step heating protocol with constant rate of $10^{12}$ K/s and identified the $T_{onset}$ by estimation of the endothermic peak of specific heat (see **Methods 2.3**).

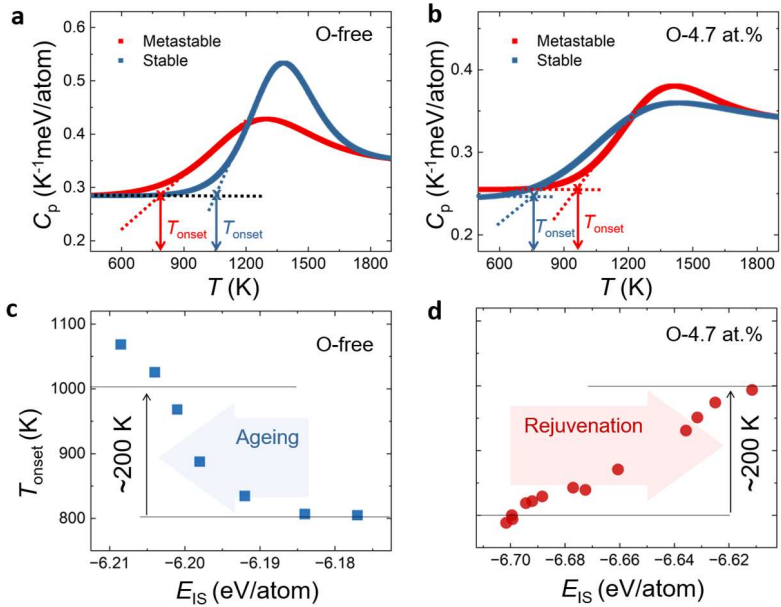

**Figure 2 Decoupling kinetic stability from thermodynamic stabilization.** The upward-scanning simulations at a heating rate of $10^{12}$ K/s for **a** O-free and **b** O-4.7 at. % glasses with different energy states. Red lines represent the temperature dependence of specific heat capacity for energetically metastable samples, and blue lines for stable ones. The correlation between the onset temperature of glass transition ($T_{onset}$) and the potential energy of inherent structures ($E_{IS}$) also exists, but with distinct dependencies for the **c** O-free and **d** O-4.7 at. % glasses respectively. A common tunable $T_{onset}$ range with a width of approximately 200 K has been identified. However, the corresponding protocols (e.g., ageing and rejuvenation) are divergent, thereby leaving room to decouple kinetic stability from thermal stability.

As additional predictions our MLIP model informed, the oxygen-free samples (Figure 2a) present characteristic enthalpy behaviors such as the thermodynamic dependence of overshoot[35]. However, as depicted in Figure 2b, the energetically stable configuration ($E_{\mathrm{IS}} = -6.702$ eV/atom) exhibits a weak resistance on thermal perturbance with temperature increasing up to 788 K, whereas the metastable one ($E_{\mathrm{IS}} = -6.677$ eV/atom) still holds the glassy state until 973 K. The deferred onset of the glass transition challenges the common belief that a lower IS energy state is indispensable to a kinetically stabilized glass[10,14,24,36]. Notably, we observe a capacity gap between these different OPSs. It is reasonable that the stable OPSs possess more constraints by elastic rigidity than that in the metastable counterparts (see SI Figure 3), leading to less freedom in vibration and a gap of heat capacity in the solid regime.

Such kinetic anomaly can be clearly exploit by the correlation with thermal profiles. Figure 2c demonstrates that the kinetic stability of oxygen-free configurations show a negative dependence of IS energy states, in consistent with some previous reports on the hyper-aged amber[10] and ultrastable metallic glasses[14]. In contrast, the hyper-quenched O-4.7 at.% samples (see Figure 2d) exhibit a positive dependence yielding $T_{\mathrm{onset}} \propto E_{IS}$. Notably, increasing the doping concentration significantly extends the upper bound of accessible stability, while the lower bound shows a modest degradation from the O-free reference (see SI Figure 4). Meanwhile, the O-4.7 at.% and O-free samples exhibit a roughly comparable range of tunable kinetic stability, serving as a gauge for mechanical performance against stability variations. Figure 3a and 3b, respectively, illustrate the mechanical performance of O-free and O-4.7 at% samples under distinct kinetic stabilities. Notably, with an equivalent enhancement in kinetic stability with $T_{\mathrm{onset}}$ increasing from 800K to 1000K, the O-4.7 at% sample maintains superior plasticity while the O-free counterpart exhibits brittleness and fracture. The diverging mechanical performance is attributed to the degree of strain localization, which can be characterized by the deviation from homogenous responses (see **Methods 2.3**). As shown in Figure 3c, the energetically favorable O-free sample has superior kinetic stability ($T_{\mathrm{onset}}$ up to 1000 K) but inferior strain uniformity ($\psi$ from 0.067 to 0.239), which is consistent with previous understanding of ultrastable glasses. However, Figure 3d presents a different scenario regarding the simultaneous preservation of kinetic stability and strain delocalization ($\psi$ from 0.082 to 0.056).

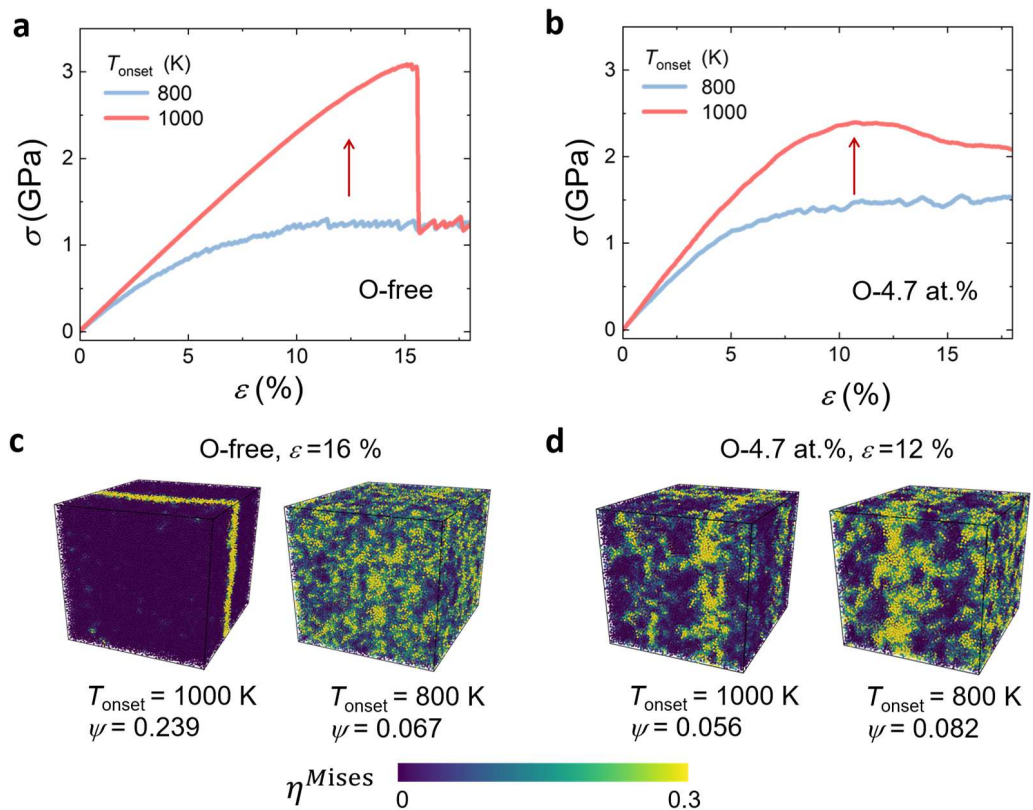

**Figure 3 Exceptional plasticity and strain delocalization by oxidation.** The strain-stress curves (**a** O-free and **b** O-4.7 at. %) and atomic mapping of von Mises strains (**c** O-free and **d** O-4.7 at. %) reveal that oxygen addition enhances ductility during elevating kinetic stability $T_{onset}$ from 800K to 1000K, whereas the remarkable enhancement cannot be sustained in oxygen-free ultrastable counterparts. $\psi$ represents the degree of strain localization: a higher value indicates a greater degree of localization, and vice versa.

**3.3 Microscopic origins and evolutional rules of pattern engineering**

Subsequently, we delve deeper into microscopic origins of the concurrent maintenance and enhancement of stability. By a stratification method[37], slowdown in sandwich confinement can be detected. However, as depicted in Figure 4a, OPSs are more geometrically complicated than sandwich walls, yielding ambiguity on the definition of distance from pinned structures. A recent work evidences that the shortest distance (red arrow) of the oxygen-target pair appears to be a streamlined descriptor of dynamical properties[18]. The description encompasses the local environments of both fluid metals and sluggish oxides. Figure 4b and 4c illustrate the pronounced slow dynamics in the OPS region with $T_{\text{onset}}$ value of 800K and 1000K (referred to as OPS1 and OPS2), respectively. The classical two-step relaxation appears with a prolonged plateau, coinciding with our previous discussions on the pinning effect of oxygen atoms in Figure 1a. Notable difference emerges when examining the metal region (Figure 4d). OPS1 abruptly terminates the decay process ($\tau_{\alpha,\text{end}} = 16.8$ ps). In contrast, OPS2 exhibits a gradual decay of slowdown until reaching a plateau value ($\tau_{\alpha,\text{end}} = \tau_{\alpha,\text{bulk}} = 3.6$ ps). Thereby two characteristic lengths were chosen to differentiate the pinned and slowdowned regions. Specifically, the correlation length of pinning, $d_0$, is identified at the temperature-independent crossover of decay curves (see SI Figure 5), reflecting the bond-dominant pinning range of the -O-Zr-O- unit[19,38] (~ 3.75 Å according to O-O pair correlation). In terms of the configuration-governed slow-down[37], we fit a universal relation $\log \left(\frac{\tau_{\alpha,d}}{\tau_{\alpha,\text{bulk}}}\right) \sim -\frac{d}{d_1 - d_0}$ to determine the value of $d_1$. The color coding in Figure 4e illustrates that OPS1 involves fewer particles in the slowdown regime compared to OPS2 in Figure 4f. Conceptually, in highly doped metallic glasses, homogeneous patches are more effective than coarsening clusters at populating the unpinned region with slowdown atoms, thereby superposing the intensity of pinning effects and suppressing the onset of the glass transition.

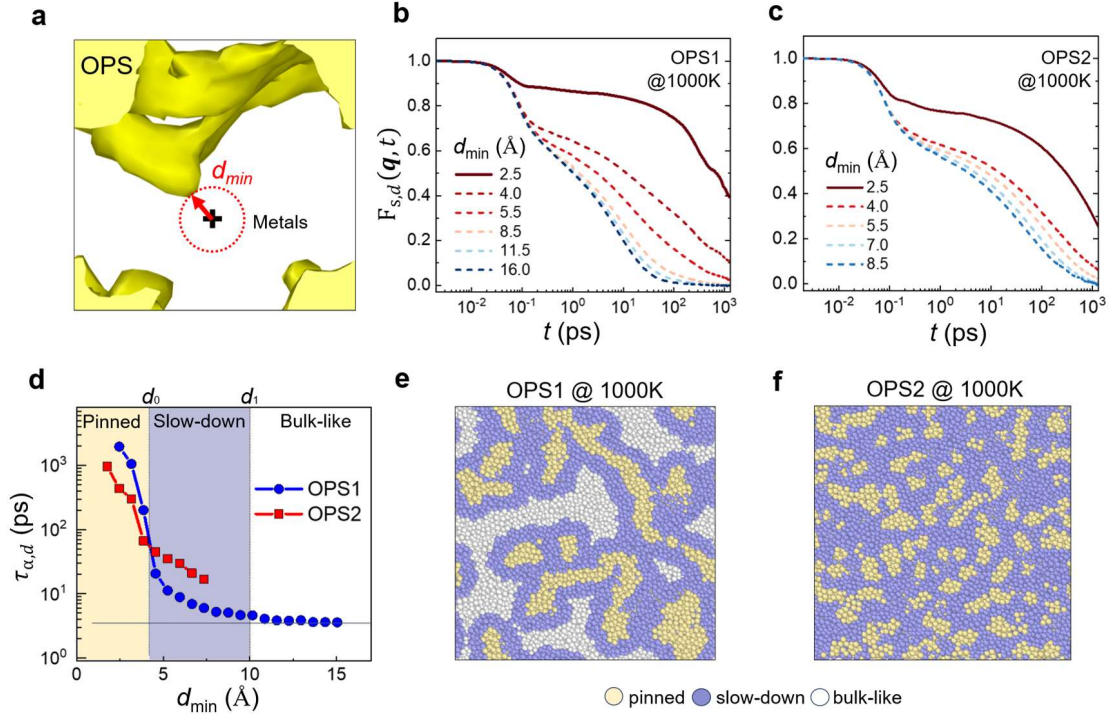

**Figure 4 Microscopic origin of exceptional kinetic stability in doped glasses. a** Schematics illustrating stratification in metals near the OPS. The OPS is visualized with yellow smooth meshes, while the remaining atoms in the metal regions are faded—except for a target atom marked by a black cross. The red circle gauges candidate target-OPS pairs, and the shortest arrow is plotted in red and refered to as '$d_{min}$'. Afterwards, the generalized incoherent intermediate scattering functions are evaluated at different distances $d_{min}$ to **b** OPS1 ($T_{onset} = 800$ K) and **c** OPS2 ($T_{onset} = 1000$ K) respectively. **d** The $d$-dependence of the α-relaxation time estimated from $F_{s,d}(q, \tau_{\alpha,d}) = 1/e$. The distance is categorized into three regimes by two correlation lengths ($d_0$ and $d_1$). $d_0$ is at the $T$-independent crossover of the stable/metastable curves. $d_1$ is fitted by logarithm of $\tau_{\alpha,d}/\tau_{\alpha,bulk}$, where $\tau_{\alpha,bulk}$ is approximated by the plateau value of $\tau_{\alpha,d}$ at $d_{min} = 15$ Å. The atomic stratification and color mapping of the pinned (yellow), slow-down (purple), and bulk-like (white) regions of **e** OPS1 and **f** OPS2 are also shown respectively.

Interestingly, geometrically ultrastable state was also computationally predicted to be not energy-favored in oxygen-free metallic glasses[39]. However, it is hard to clarify the structural mechanism due to the "black box" nature of the artificial intelligence like graph neural networks. Although we have partially addressed this issue by the OPS-assisted stratification, the lack of a sound interpretable framework to quantify the structural evolution is all the more detrimental. Namely, while the kinetic stability can be predicted by the as-sampled glassy structures, the routine of inverse design to those unique states remains unclear. Motivated by the dramatical slowdown (see Figure 1a), in conjunction with strong affinity between oxygen and zirconium atoms[38], a direct inference can be drawn that atoms in the oxidized metallic glasses manifest distinct diffusion behaviors (see Figure 5a). The discrepancy in mobility of inhibitors and activators is a pre-requisite of the self-organized pattern formation in the reaction-diffusion model proposed by Turing[40]. However, in Figure 5b, the short-range and medium-range connection degrees indicate that oxygen-centered clusters tend to interconnect but rarely conform to stoichiometry, dictating insufficient feedback flux from slow atoms to fast atoms. As a result, the local fluctuation of OPS, which functions similarly to autocatalysis in Turing dynamics, cannot be compensated, causing the patterns to continuously coarsen[41] (also disclosed in Figure 1e). As an indicator of coarsening, the size distribution of the oxidation-induced patches is anticipated to exhibit a shift in peak position from metastable to stable states (shown in SI Figure 6), with an average radius increasing from 5.6 to 9.4 Å (see **Methods 2.5** for the detailed definition of cluster radius).

Inspired by the aggregation dynamics in the patchy patterned ecosystem[42], we mimic the energy-favored polymorphism of OPSs in a generalized reaction-diffusion equations

$$\frac{\partial u}{\partial t} = f(u,v) + \delta \nabla^2 u \qquad (1a)$$

$$\frac{\partial v}{\partial t} = -f(u,v) + \nabla^2 v \qquad (1b)$$

where $f(u,v) = uv - \frac{\alpha u}{u+\beta}$ and $\delta$ represents the ratio of diffusion rates $\frac{D_u}{D_v}$. We adjust the reaction term of original Turing equations to enforce a conservation form, which is an intrinsic requirement in atomistic simulations. In our model, the activator $u$ and inhibitor $v$ represent a local concentration of particles in less mobile and highly mobile form, respectively. For clarity, we choose the mean squared displacement of the oxygen-free metallic glasses as a mobility classifier (see Figure 5a). Estimation of the values of parameters $\alpha$, $\beta$, and $\delta$ and the initial condition of $u$ and $v$ can be found in

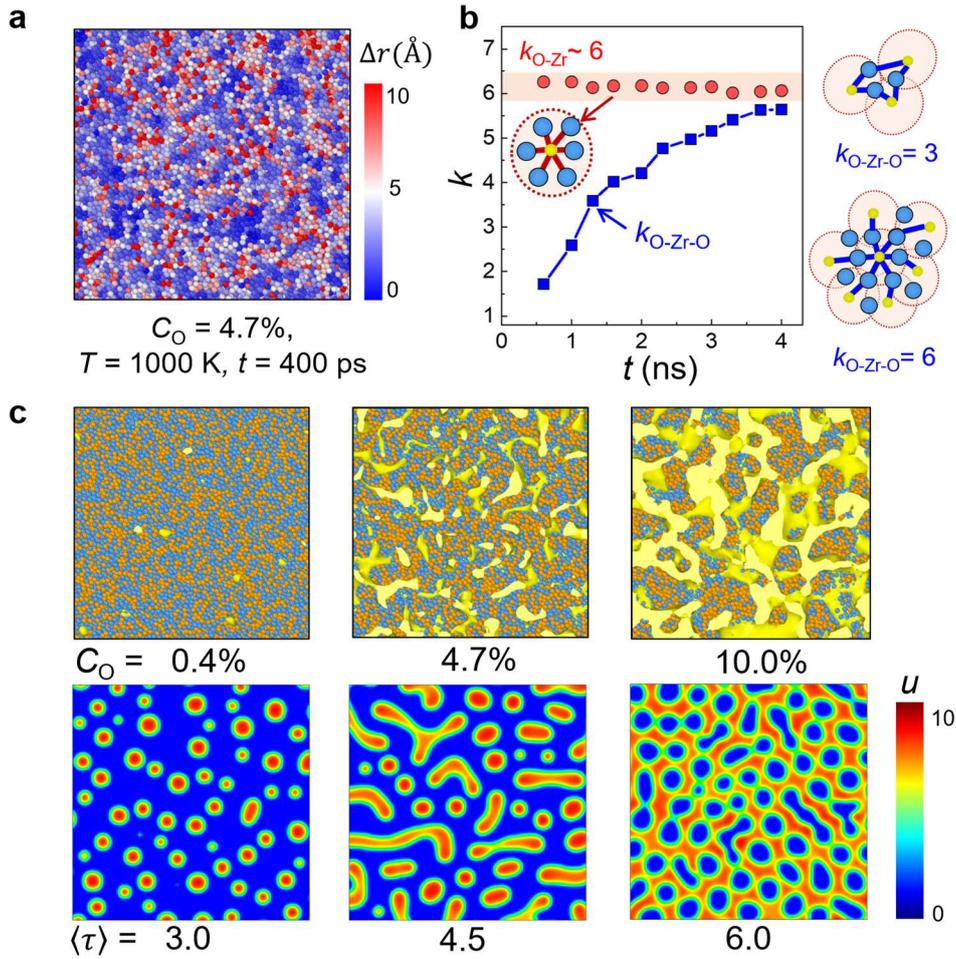

**Figure 5 Formation mechanism of OPS patterns. a** Classification of fast and slow particles in a homogenous OPS in the O-4.7 at. % sample. The mean squared root of displacement of the oxygen-free MG (~ 5 Å at 1000 K) after 400 ps was used as a criterion. In terms of the reaction-diffusion model, inhibitors (fast particles) and activators (slow particles) are pre-requisites for patterns. **b** The degree of connections within a single cluster ($k_{O-Zr}$) and across the entire network ($k_{O-Zr-O}$) is presented. The bonding network is simplified into a 2D ball-and-stick diagram, where yellow balls represent central oxygen atoms and blue ones zirconium atoms. **c** Stationary patterns in OPSs (upper panel) and numerical snapshots (lower panel) with varying density. OPSs were sampled from a 4-ns canonical MD at 1500 K, and numerical snapshots were generated after 90000 timesteps of reaction-diffusion calculations.

**Methods 2.4**. For mathematical considerations to ensure the spinodal decomposition, we normalized the initial value of $u$ and $v$ by specifying the global packing density $\langle \tau \rangle \equiv \langle u + v \rangle$ in a pattern-allowed range. The bracket is related to a spatial average. To examine the validity of the refined equations, we conducted canonical MD simulations on OPS pattern evolution at 1500 K. The temperature is much higher than $T_{\text{onset}}$ but still not enough for a further upheaval on OPSs, implying a large diffusion ratio which facilitates the pattern formation. Atomistic and numerical snapshots after thorough sampling are shown in Figure 5c. Patterns in the numerical models and MD-based samples possess similar characteristic structures, which vary from isolated clumps, gapped stripe, to a percolated network as the density of oxygen increases. From the notable local density fluctuations between pixels, we can also unambiguously define the oxide and metal phases by the abundance of the slow particles. Clearly, the oxide phase completely covers the OPS region, leaving the metal phase to accommodate for the remaining metallic atoms. This phenomenon known as phase separation was confirmed very recently on the oxidized metallic glass nanotubes[19] and nanomembranes[18]. Our model is also capable for minor alloying (< 1000 atom ppm) and severed oxidized (> 30 at. %) contents which generally behaves like a homogenous phase with a $\langle\tau\rangle$ out of the Turing-Hopf range[43].

**3.4 Masking strategy for atomistic reconstructions of pattern engineering**

Thus far, validation has been limited to the qualitative analysis of the steady-state solution. Further investigations into coarsening processes necessitate a quantitative pattern-mapping approach, which deliberately translates the discrete pixels from numerical results into atomistic samples for MD simulations. Our pattern-masking strategy is illustrated in Figure 6a (also see **Methods 2.5**). The patterned OPS closely resembles the mask, maintaining an almost identical spatial distribution of patches in MD samples. Figure 6b delineates the relationship between kinetic stability and inherent energy states. Patterns generated from direct engineering collapses to the same linear relation gauged by the REMD-sampled counterparts. The results arguably reinforce our conclusions on the metastable OPSs and additionally provide a new routine to prepare structurally stable states without multifaceted thermal protocols. Moreover, the patterned OPSs fall in relatively high energy basins, approaching far-from-equilibrium states which are rarely sampled even through enhanced sampling.

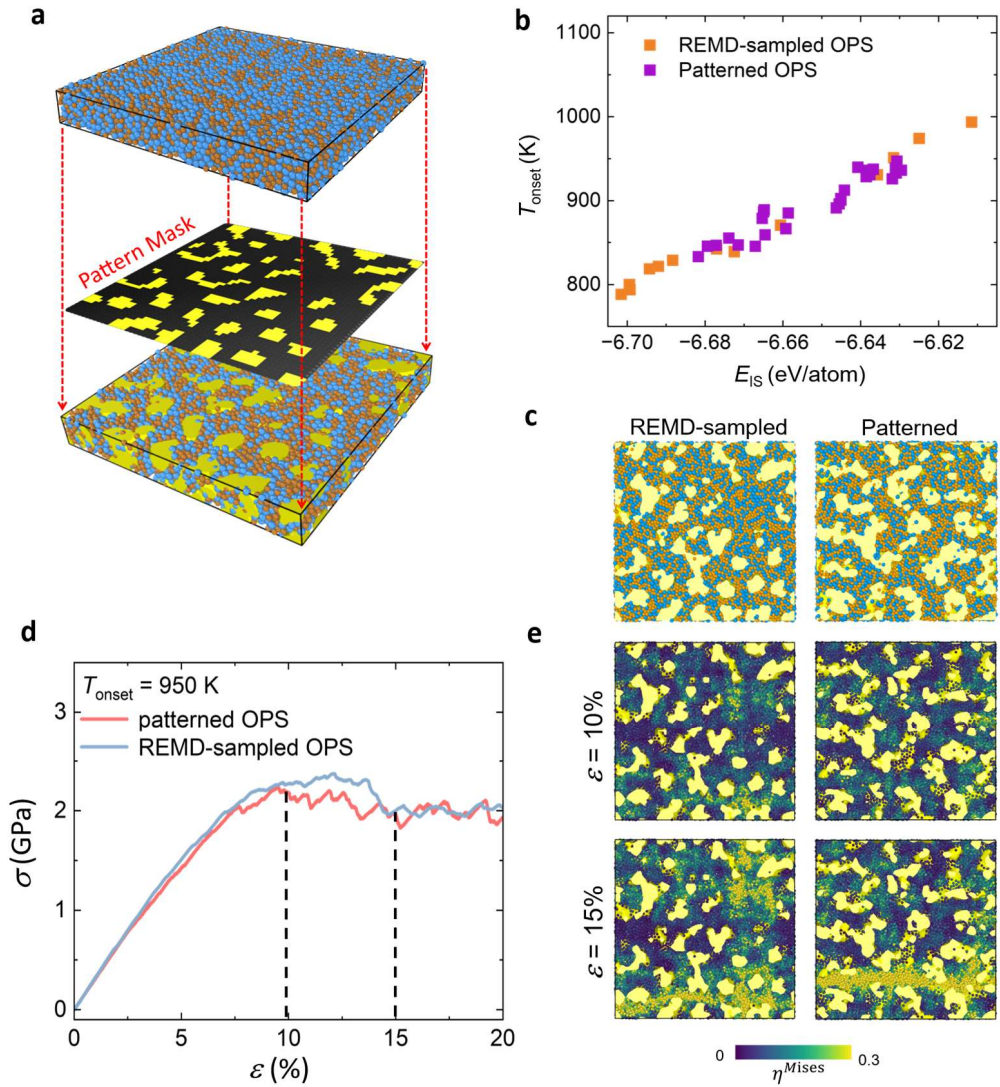

**Figure 6 Pattern engineering of kinetic and mechanical performance. a** Schematics of the pattern-masked mapping. The black mesh grids on the pattern mask are filled with yellow pixels, which correspond to clusters or patches generated by the reaction-diffusion model. **b** Kinetic stability of REMD-sampled (orange) and patterned (purple) OPS. **c** Profiles of the REMD-sampled (orange) and the patterned (purple) OPS with kinetic hyper-stability. **d** Strain-stress curves and **e** shear snapshots of the two OPSs are also presented. The yellow surface meshes in panels **c** and **e** represent the geometry and distribution of OPSs.

We also demonstrate that pattern engineering can recurrent the REMD-sampled patches, not only in stationary states (Figure 6c) but also in coarsening processes with an average radius increasing from 4.861 to 6.808 Å (SI Figure 7). We further examined coarsening behaviors within the Cahn-Hillard theoretical frameworks[44]. Given the Ostward ripening approxiamation[45], the minority phase forms spherical droplets, and the smaller droplets are absorbed through diffusion into the larger ones. Lifshitz and Slyozov[46] and Wagner[47] have derived a probability density function for those droplets in quasi-equilibrium. Neither the REMD-sampled (SI Figure 6) nor the patterned distributions (SI Figure 7) align with the theoretical diffusion-dominant predictions, indicating the reaction-dominant behaviors, such as pinning effects and Zr-O bonding, play a crucial role in pattern formation. The protocol-free strategy can also effectively maintain favorable mechanical performance. Observations from Figure 6d reveal that *patterned OPS with high kinetic stability share a striking consistency with OPS samples obtained via REMD sampling*, with mechanical responses largely unimpaired. The dynamic process underlying this phenomenon is further visualized through key frames captured during shear loading (Figure 6e), where both REMD-sampled and patterned OPSs can induce forced branching of shear bands, thereby preserving remarkable strain delocalization capability. This mechanism of toughening and strengthening through shear band branching aligns closely with experimental results[30], providing direct evidence for elucidating the relevant phenomena.

**Discussion**

In summary, we demonstrate that doped metallic glasses can be energetically metastable yet kinetically stable by pattern engineering, which paves a new way to stabilize the disordered structures of amorphous materials and advances understanding of structural mechanism on dual phase nanostructures[48,49]. In a microscopic view, oxygen contents in the amorphous metallic environment exhibit extremely slow diffusion, leading to collectively sluggish dynamics[50] and locally dense packing[30]. Motivated by these observations, we infer that tailored oxygen dopants might serve as a critical proxy to realize pinning in bulk[32], ribbon[51], and low-dimensional[18,19] MGs. Given that the pinning strength evolves differently under various protocols (such as distinct heating rates and/or annealling temperatures), doping and dispersing strategies beyond traditional casting must be meticulously designed to achieve the targeted OPS.

This explains why the high-performance OPS is typically observed in low-dimensional metals rather than in bulk-like ones. We expect that the formalism can be carried out of quasi-two-dimensional plate-like configurations back to three-dimensional structures using layer-by-layer reconstructions.

This work also challenges the conventional notion that kinetically hyperstable glasses must occupy low-energy basins within the potential energy landscape (PEL). By demonstrating that kinetic stability can be selectively enhanced through dopant patterning, we establish a framework for designing functional amorphous materials without sacrificing ductility. With assistance from machine-learning interatomic potentials, this framework can be straightforwardly implemented on other prototypical alloys. By controlling the intensity of reaction-like catalysis, such as doping concentrations and spatial distribution of impurity, we can engineer the morphology of patterned regions and fabricate materials with desired functionality. Our findings not only resolve the longstanding stability-ductility trade-off but also open avenues for digital twinning in nanostructured alloys, additive manufacturing, and high-strength composites, where both resistance to devitrification and mechanical resilience are critical.

## CRediT authorship contribution statement

**Huanrong Liu**: Conceptualization, Data curation, Formal analysis, Software, Methodology, Visualization, Writing – original draft, Writing - review & editing. **Qingan Li**: Writing - review & editing, Software. **Shan Zhang**: Writing - review & editing, Supervision. **Rui Su**: Writing - review & editing, Software. **Yunjiang Wang**: Writing - review & editing, Validation, Supervision, Conceptualization. **Pengfei Guan**: Writing - review & editing, Validation, Supervision, Conceptualization, Funding acquisition, Resources.

## Declaration of interests

The authors declare that they have no known competing financial interests or personal relationships that could have appeared to influence the work reported in this paper.

## Acknowledgements

Y.W. was supported by the Strategic Priority Research Program of Chinese Academy of Sciences (grants nos. XDB0620103 and XDB0510301), and the National Natural Science Foundation of China (grant no. 12472112). R.S. acknowledges the Young Scientists Fund of the National Natural Science Foundation of China (51801046). P.G. and R.S. acknowledge the support from Advanced Materials-National Science and Technology Major Project (No. 2024ZD0606900). P.G. was also supported by the National Natural Science Foundation of China (Nos. 52161160330 and T2325004). We also thank the computational support from Beijing Computational Science Research Center (CSRC).

# References

1 Kauzmann, W. The Nature of the Glassy State and the Behavior of Liquids at Low Temperatures. *Chemical Reviews* **43**, 219-256, doi:10.1021/cr60135a002 (1948).

2 Liu, X., Queen, D. R., Metcalf, T. H., Karel, J. E. & Hellman, F. Hydrogen-Free Amorphous Silicon with No Tunneling States. *Physical Review Letters* **113**, 025503, doi:10.1103/PhysRevLett.113.025503 (2014).

3 Pérez-Castañeda, T., Rodríguez-Tinoco, C., Rodríguez-Viejo, J. & Ramos, M. A. Suppression of tunneling two-level systems in ultrastable glasses of indomethacin. *Proceedings of the National Academy of Sciences* **111**, 11275-11280, doi:10.1073/pnas.1405545111 (2014).

4 Wang, Y., Qian, Z., Tong, H. & Tanaka, H. Hyperuniform disordered solids with crystal-like stability. *Nature Communications* **16**, 1398, doi:10.1038/s41467-025-56283-1 (2025).

5 Ozawa, M., Iwashita, Y., Kob, W. & Zamponi, F. Creating bulk ultrastable glasses by random particle bonding. *Nature Communications* **14**, 113, doi:10.1038/s41467-023-35812-w (2023).

6 Swallen, S. F. *et al.* Organic Glasses with Exceptional Thermodynamic and Kinetic Stability. *Science* **315**, 353-356, doi:10.1126/science.1135795 (2007).

7 Singh, S., Ediger, M. D. & de Pablo, J. J. Ultrastable glasses from in silico vapour deposition. *Nature Materials* **12**, 139-144, doi:10.1038/nmat3521 (2013).

8 Ninarello, A., Berthier, L. & Coslovich, D. Models and Algorithms for the Next Generation of Glass Transition Studies. *Physical Review X* **7**, 021039, doi:10.1103/PhysRevX.7.021039 (2017).

9 Swallen, S. F. *et al.* Organic Glasses with Exceptional Thermodynamic and Kinetic Stability. **315**, 353-356, doi:doi:10.1126/science.1135795 (2007).

10 Pérez-Castañeda, T., Jiménez-Riobóo, R. J. & Ramos, M. A. Two-Level Systems and Boson Peak Remain Stable in 110-Million-Year-Old Amber Glass. *Physical Review Letters* **112**, 165901, doi:10.1103/PhysRevLett.112.165901 (2014).

11 Kob, W. & Berthier, L. Probing a liquid to glass transition in equilibrium. *Phys Rev Lett* **110**, 245702, doi:10.1103/PhysRevLett.110.245702 (2013).

12 Yu, H.-B., Luo, Y. & Samwer, K. Ultrastable Metallic Glass. *Advanced Materials* **25**, 5904-5908, doi:https://doi.org/10.1002/adma.201302700 (2013).

13 Luo, P. *et al.* Ultrastable metallic glasses formed on cold substrates. *Nature Communications* **9**, 1389, doi:10.1038/s41467-018-03656-4 (2018).

14 Zhao, Y. *et al.* Ultrastable metallic glass by room temperature aging. *Science Advances* **8**, eabn3623, doi:10.1126/sciadv.abn3623.

15 Louzguine-Luzgin, D. V. & Jiang, J. On Long-Term Stability of Metallic Glasses. *Metals* **9** (2019).

16 Zeng, Q.-s. *et al.* Origin of Pressure-Induced Polyamorphism in Ce75Al25 Metallic Glass. *Physical Review Letters* **104**, 105702, doi:10.1103/PhysRevLett.104.105702 (2010).

17 Bosoni, E. *et al.* How to verify the precision of density-functional-theory implementations via reproducible and universal workflows. *Nature Reviews Physics* **6**, 45-58, doi:10.1038/s42254-023-00655-3 (2024).

18 Yang, Z. *et al.* Oxide-Metal Hybrid Glass Nanomembranes with Exceptional Thermal Stability. *Nano Letters* **24**, 14475-14483, doi:10.1021/acs.nanolett.4c04555 (2024).

19 Li, F. *et al.* Oxidation-induced superelasticity in metallic glass nanotubes. *Nat Mater* **23**, 52-57, doi:10.1038/s41563-023-01733-8 (2024).

20  Dong, Z., Ma, Z., Yu, L. & Liu, Y. Achieving high strength and ductility in ODS-W alloy by employing oxide@W core-shell nanopowder as precursor. *Nature Communications* **12**, 5052, doi:10.1038/s41467-021-25283-2 (2021).

21  Parrinello, M. & Rahman, A. Polymorphic transitions in single crystals: A new molecular dynamics method. *Journal of Applied Physics* **52**, 7182-7190, doi:10.1063/1.328693 %J Journal of Applied Physics (1981).

22  Fan, Y., Iwashita, T. & Egami, T. Energy landscape-driven non-equilibrium evolution of inherent structure in disordered material. *Nat Commun* **8**, 15417, doi:10.1038/ncomms15417 (2017).

23  Thompson, A. P. *et al.* LAMMPS - a flexible simulation tool for particle-based materials modeling at the atomic, meso, and continuum scales. *Computer Physics Communications* **271**, 108171, doi:https://doi.org/10.1016/j.cpc.2021.108171 (2022).

24  Wang, L.-M., Velikov, V. & Angell, C. A. Direct determination of kinetic fragility indices of glassforming liquids by differential scanning calorimetry: Kinetic versus thermodynamic fragilities. *The Journal of Chemical Physics* **117**, 10184-10192, doi:10.1063/1.1517607 %J The Journal of Chemical Physics (2002).

25  Mirdamadi, E. S., Haghbin Nazarpak, M. & Solati-Hashjin, M. in *Structural Biomaterials* (ed Cuie Wen) 301-331 (Woodhead Publishing, 2021).

26  Cheng, Y. Q., Cao, A. J. & Ma, E. Correlation between the elastic modulus and the intrinsic plastic behavior of metallic glasses: The roles of atomic configuration and alloy composition. *Acta Materialia* **57**, 3253-3267, doi:https://doi.org/10.1016/j.actamat.2009.03.027 (2009).

27  Rycroft, C. H. VORO++: A three-dimensional Voronoi cell library in C++. *Chaos: An Interdisciplinary Journal of Nonlinear Science* **19**, doi:10.1063/1.3215722 (2009).

28  Wang, J. *et al.* Clustering-mediated enhancement of glass-forming ability and plasticity in oxygen-minor-alloyed Zr-Cu metallic glasses. *Acta Materialia* **261**, 119386, doi:https://doi.org/10.1016/j.actamat.2023.119386 (2023).

29  Stillinger, F. H. A Topographic View of Supercooled Liquids and Glass Formation. *Science* **267**, 1935-1939, doi:10.1126/science.267.5206.1935 (1995).

30  Wu, Y. *et al.* Substantially enhanced plasticity of bulk metallic glasses by densifying local atomic packing. *Nat Commun* **12**, 6582, doi:10.1038/s41467-021-26858-9 (2021).

31  Zella, L., Moon, J. & Egami, T. Ripples in the bottom of the potential energy landscape of metallic glass. *Nat Commun* **15**, 1358, doi:10.1038/s41467-024-45640-1 (2024).

32  Zhou, W. H. *et al.* Effect of alloying oxygen on the microstructure and mechanical properties of Zr-based bulk metallic glass. *Acta Materialia* **220**, 117345, doi:https://doi.org/10.1016/j.actamat.2021.117345 (2021).

33  Lyubartsev, A. P., Martsinovski, A. A., Shevkunov, S. V. & Vorontsov‐Velyaminov, P. N. New approach to Monte Carlo calculation of the free energy: Method of expanded ensembles. *The Journal of Chemical Physics* **96**, 1776-1783, doi:10.1063/1.462133 %J The Journal of Chemical Physics (1992).

34  Lin, X. H., Johnson, W. L. & Rhim, W. K. Effect of Oxygen Impurity on Crystallization of an Undercooled Bulk Glass Forming Zr-Ti-Cu-Ni-Al Alloy. *Materials Transactions, JIM* **38**, 473-477, doi:10.2320/matertrans1989.38.473 (1997).

35   Chen, Z. *et al.* Dependence of calorimetric glass transition profiles on relaxation dynamics in non-polymeric glass formers. *Journal of Non-Crystalline Solids* **433**, 20-27, doi:10.1016/j.jnoncrysol.2015.11.021 (2016).

36   Zhang, F. *et al.* Effects of sub-Tg annealing on Cu64.5Zr35.5 glasses: A molecular dynamics study. *Applied Physics Letters* **104**, doi:10.1063/1.4864652 (2014).

37   Hu, Y.-C. *et al.* Configuration correlation governs slow dynamics of supercooled metallic liquids. *Proceedings of the National Academy of Sciences* **115**, 6375-6380, doi:10.1073/pnas.1802300115 (2018).

38   Wang, Z. *et al.* Effects of Oxygen Impurities on Glass-Formation Ability in Zr2Cu Alloy. *The Journal of Physical Chemistry B* **120**, 9223-9229, doi:10.1021/acs.jpcb.6b06306 (2016).

39   Wang, Q. & Zhang, L. Inverse design of glass structure with deep graph neural networks. *Nat Commun* **12**, 5359, doi:10.1038/s41467-021-25490-x (2021).

40   Turing, A. M. The chemical basis of morphogenesis. *Philosophical Transactions of the Royal Society of London. Series B, Biological Sciences* **237**, 37-72, doi:10.1098/rstb.1952.0012 (1957).

41   Liu, Q. X. *et al.* Phase separation explains a new class of self-organized spatial patterns in ecological systems. *Proc Natl Acad Sci U S A* **110**, 11905-11910, doi:10.1073/pnas.1222339110 (2013).

42   Siteur, K. *et al.* Phase-separation physics underlies new theory for the resilience of patchy ecosystems. *Proc Natl Acad Sci U S A* **120**, e2202683120, doi:10.1073/pnas.2202683120 (2023).

43   Lengyel, I. & Epstein, I. R. Modeling of Turing Structures in the Chlorite—Iodide—Malonic Acid—Starch Reaction System.  **251**, 650-652, doi:doi:10.1126/science.251.4994.650 (1991).

44   Cahn, J. W. & Hilliard, J. E. Free Energy of a Nonuniform System. I. Interfacial Free Energy. *The Journal of Chemical Physics* **28**, 258-267, doi:10.1063/1.1744102 %J The Journal of Chemical Physics (1958).

45   Baldan, A. Review Progress in Ostwald ripening theories and their applications to nickel-base superalloys Part I: Ostwald ripening theories. *Journal of Materials Science* **37**, 2171-2202, doi:10.1023/A:1015388912729 (2002).

46   Lifshitz, I. M. & Slyozov, V. V. The kinetics of precipitation from supersaturated solid solutions. *Journal of Physics and Chemistry of Solids* **19**, 35-50, doi:https://doi.org/10.1016/0022-3697(61)90054-3 (1961).

47   Wagner, C. Theorie der Alterung von Niederschlägen durch Umlösen (Ostwald-Reifung). *Zeitschrift für Elektrochemie, Berichte der Bunsengesellschaft für physikalische Chemie* **65**, 581-591, doi:https://doi.org/10.1002/bbpc.19610650704 (1961).

48   Wu, G., Chan, K. C., Zhu, L., Sun, L. & Lu, J. Dual-phase nanostructuring as a route to high-strength magnesium alloys. *Nature* **545**, 80-83, doi:10.1038/nature21691 (2017).

49   Wu, G. *et al.* Hierarchical nanostructured aluminum alloy with ultrahigh strength and large plasticity. *Nat Commun* **10**, 5099, doi:10.1038/s41467-019-13087-4 (2019).

50   Damodaran, K. V., Nagarajan, V. S. & Rao, K. J. A molecular dynamics study of ZrO2     SiO2 system. *Journal of Non-Crystalline Solids* **124**, 233-241, doi:https://doi.org/10.1016/0022-3093(90)90268-Q (1990).

51      Sun, S., Chai, T., Yao, K., Zhang, Z. & Chen, N. New Class of Amorphous Oxide Semiconductors from Amorphous Alloys. *The Journal of Physical Chemistry C* **127**, 24411-24416, doi:10.1021/acs.jpcc.3c05958 (2023).

# Supplementary Information - Turing Pattern Engineering Enables Kinetically Ultrastable yet Ductile Metallic Glasses

Huanrong Liu[1], Qingan Li[1], Shan Zhang[2,3], Rui Su[4,*], Yunjiang Wang[5,6,*]，Pengfei Guan[2,3,1,*]

[1] Beijing Computational Science Research Center, Beijing 100193, China.

[2] Advanced Interdisciplinary Science Research (AiR) Center, Ningbo Institute of Materials Technology and Engineering, Chinese Academy of Sciences, Ningbo 315201, China.

[3] State Key Laboratory of Advanced Marine Materials, Ningbo Institute of Materials Technology and Engineering, Chinese Academy of Sciences, Ningbo 315201, China.

[4] College of Materials & Environmental Engineering, Hangzhou Dianzi University, Hangzhou 310018, China.

[5] State Key Laboratory of Nonlinear Mechanics, Institute of Mechanics, Chinese Academy of Sciences, Beijing 100190, China.

[6] School of Engineering Science, University of Chinese Academy of Sciences, Beijing 101408, China

**\*Corresponding authors.** surui@hdu.edu.cn (R.S.), yjwang@imech.ac.cn (Y.W.), pguan@nimte.ac.cn (P.G.)

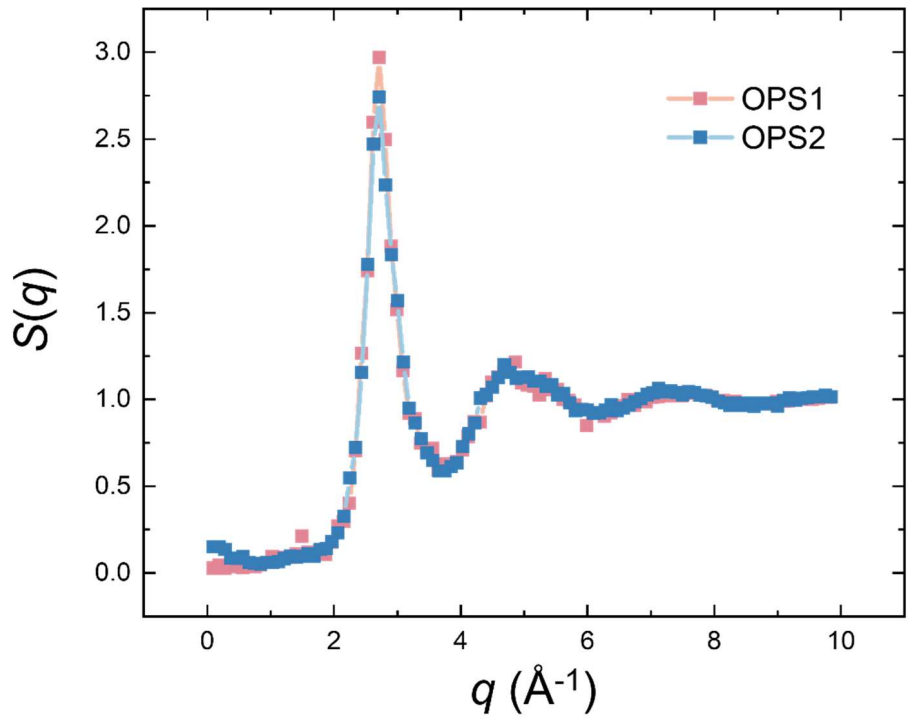

**SI Figure 1 | Structure factors of different OPSs**. Different OPSs have very close positions of the first peak, which makes it very difficult to distinguish OPSs with different morphologies from experimental results..

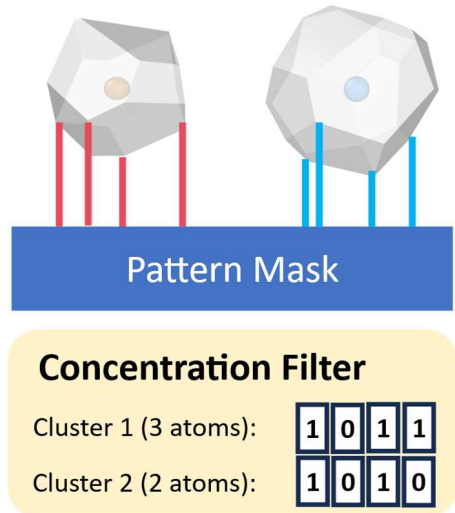

SI Figure 2 | Pattern mask preparation. A mask was generated based on the numerical spatio-temporal dynamics of the reaction-diffusion equations, while a plate-like oxygen-free metallic glass served as the reference configuration for Voronoi tessellation. The mask pixels were rescaled to match the plate size, and the vertices of the Voronoi polyhedra were grouped by the pattern mask and filtered by a pre-defined doping concentration. Interstitial oxygen atoms were subsequently inserted at selected vertices and optimized. The Voronoi vertices are collected and filtered to the candidates for insertion.

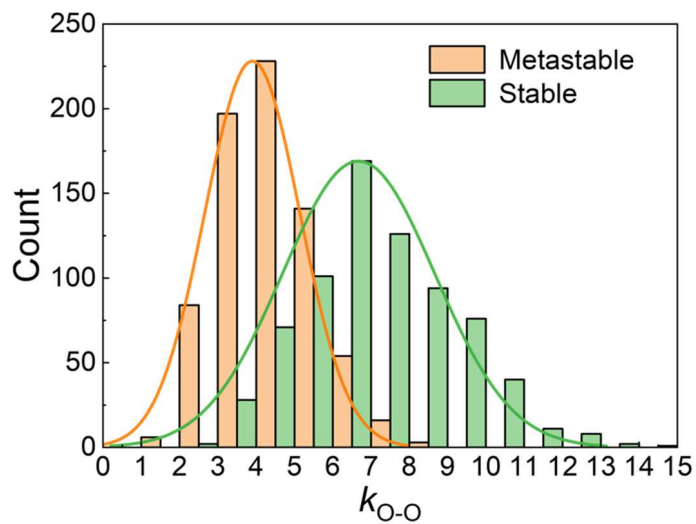

**SI Figure 3 | The O-O connectivity freedom in elastic rigidity of OPS.** It is characterized by the local connectivity of -O-Zr-O- unit within a 4.2 Å cutoff radius. The enumeration can be fitted by gaussian distribution, with obvious peak shifting from the energetically stable OPS to the metastable counterpart.

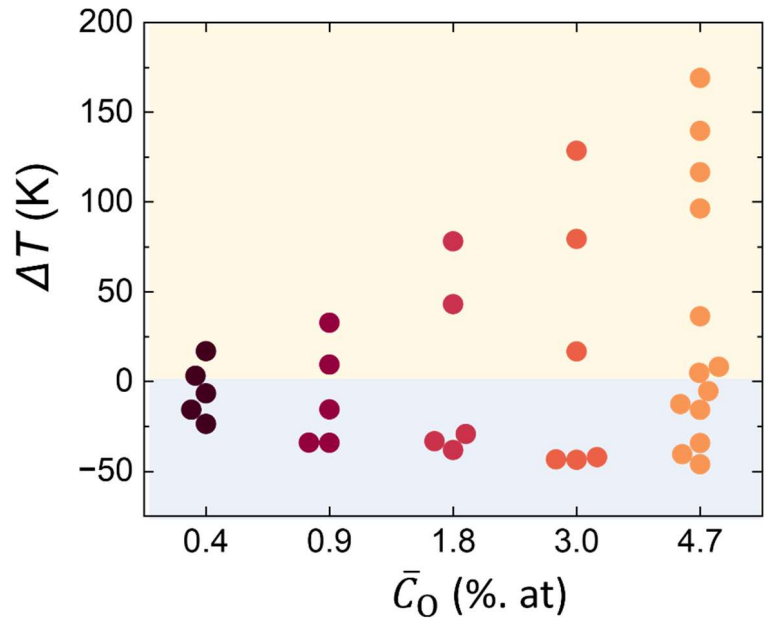

**SI Figure 4 | Propagated range of tuable onset temperature.** Bee swarm plot of the tunable onset temperature of glass transition as a function of doping concentration. The parent O-free configuration ($\Delta T = 0$) is indicated by the black bottom ($T_{\text{onset}} = 800$ K) in the Figure 2c. Such coexistence of promotion and inhibition with the same doping concentration indicates that energy state plays a crucial role on the stability tuning, which rationalizes the sometimes contradictory observations in experiments. Moreover, the propagated range indicates remarkable tunability of kinetic stability for highly doped metallic glasses. Noted that in the minor alloying regime ($\overline{C_O} \leq 0.4\%$), the infinitesimal change in $T_{\text{onset}}$ is in agreement with experiments on bulk metallic glasses.

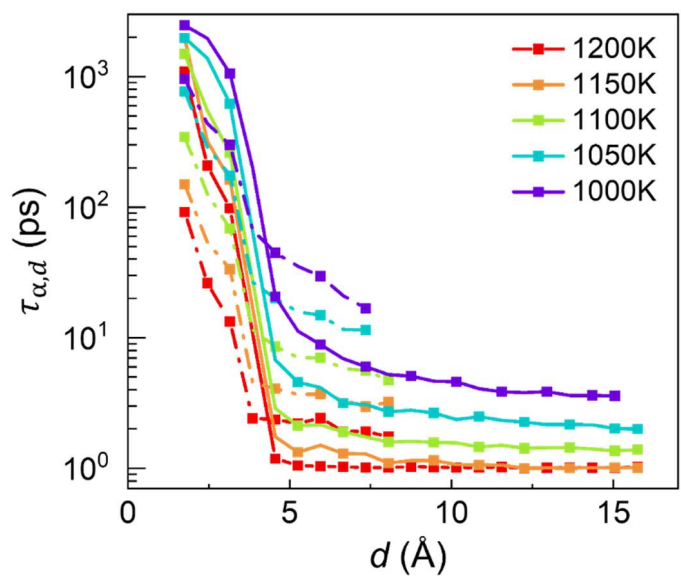

**SI Figure 5 | Decoupling of pinning correlation length with temperature.** The dynamical crossover of stable (bold lines) and metastable (dashed lines) OPS is temperature-independent, offering a well-defined characterization of pinning effects.

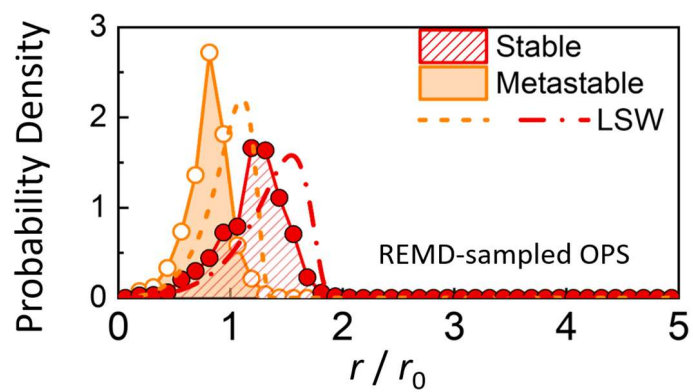

**SI Figure 6 | Patch-size distribution of REMD-sampled OPSs**. The radius is normalized by the mask pixel size ($r_0 = 5$ Å). Dashed lines represent theoretical predictions from the Lifshitz–Slyozov–Wagner (LSW) model, which is derived from the Cahn-Hillard equations in the Ostward-ripening approximation.

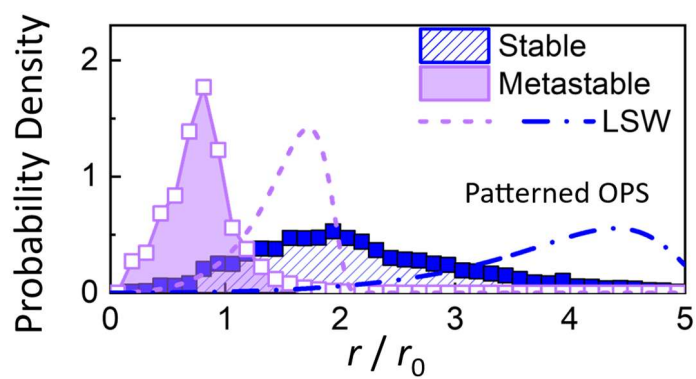

**SI Figure 7 | Patch-size distribution of patterned OPSs**. The radius is normalized by the mask pixel size ($r_0 = 5$ Å). Dashed lines represent theoretical predictions from the Lifshitz–Slyozov–Wagner (LSW) model.